\documentstyle[seceq,epsf,wrapft,twoside]{ptptex}
\notypesetlogo  %comment in if to eliminate PTPTeX logo
\title{Covariant Classification Scheme of Hadrons and Chiral States
}
\author{%
Shin {\sc Ishida}$^1$ and Muneyuki {\sc Ishida}$^2$
}
\inst{%
$^{1}$ Research Institute of Quantum Science, 
College of Science and Technology\\
Nihon University, Tokyo 101-0062, Japan\\
$^{2}$ Department of Physics, Meisei University, Hino 191-8506, Japan
}
\recdate{%
\today
}
\abst{%
The essential points and physical backgrounds of the covariant level-classification scheme,
recently proposed by us, are reviewed:
In this scheme the general composite hadrons are classified as the multiplets with the 
$\tilde U(12)_{SF}\times O(3,1)$ symmetry, where 
$\tilde U(12)_{SF}\supset \tilde U(4)_{\rm D.S.}\bigotimes SU(3)_{\rm F}$, and 
$\tilde U(4)_{\rm D.S.}$ is a homogeneous Lorentz transformation group for Dirac spinors.
This symmetry predicts the existence of new type of 
chiral mesons and baryons out of the conventional level classification scheme.
The $\sigma$ nonet is a typical example of chiralons to be assigned to the $(q\bar q)$ 
relativistic $S$-wave state. 
}
\begin{document}
\maketitle

\setcounter{tocdepth}{4}

\section{Introduction}

({\it Present status of level-classification and purpose of this talk})\ \ \ 
There exist the two contrasting, non-relativistic and relativistic, 
viewpoints of level-classification. The former is based on the 
non-relativistic quark model (NRQM) with the approximate $LS$-symmetry 
and gives a theoretical base to the PDG level-classification. The 
importance of the role of a relativistic symmetry, chiral symmetry, played in hadron physics
is widely accepted, and
it is well-known that $\pi$ meson octet has 
the property as a Nambu-Goldstone boson in the case of spontaneous 
breaking of chiral symmetry.

Owing to the recent progress, 
the existence of light $\sigma$-meson as chiral partner of $\pi (140)$
seems to be established\cite{rf1} especially through the analysis of 
various $\pi\pi$-production processes. This gives further a strong 
support to the relativistic viewpoint.

Thus, the hadron spectroscopy is now being confronted with a serious problem,
existence of the seemingly contradictory two viewpoints, Non-relativistic 
and Extremely Relativistic ones.

Recently, corresponding to the above situation, we have proposed\cite{rf2,rf3} 
a covariant level-classification  
scheme of 
light-through-heavy quarks (and possibly of gluonic) hadrons,
unifying the above two view-points. There we have pointed out a possibility 
that an approximate symmetry of 
$\tilde U(12)_{SF} \supset \tilde U(4)_{\rm D.S.} \bigotimes U(3)_{\rm F}$
is realized in nature in the world of 
hadron spectroscopy.

Here the $\tilde U(12)_{SF}$ symmetry is mathematically the same as the one\cite{rf4} 
that appeared in 1965 to generalize covariantly the static $SU(6)_{\rm SF}$ 
symmetry\cite{rf5} ($SU(6)_{\rm SF} \supset SU(2)_{\rm S}\bigotimes U(3)_{\rm F}$)
in NRQM. However, in the case of $\tilde U(12)_{\rm SF}$ at that time 
{\it only\ the\ boosted\ Pauli-spinors}(, which reduce to the Pauli-spinors at the hadron
rest frame,) are taken as 
{\it physical components} out of the fundamental representations of 
$\tilde U(4)_{\rm D.S.}$. Now in the present scheme\footnote{
We use the term $\tilde U(4)_{\rm D.S.}$ ($\tilde U(12)_{SF}$) as what implying $U(4)_{\rm stat.}$
 ($U(12)_{\rm stat.}$), that is, 4(12)-dimensional unitary group at the rest-frame.
} 
of the $\tilde U(12)$ 
symmetry {\it all\ general\ Dirac\ spinors} are, inside of hadrons, 
to be treated as {\it physical}.

Accordingly in the covariant classification scheme 
there exist\footnote{
Shortly after the symposium, the existence of new narrow resonances, $D_{s,J}(2317)$ and $D_{s,J}(2463)$
in $(c\bar s)$ system, aroused strong interests among us. They are naturally assigned\cite{rf19} 
as the ground state
chiralons with $J^P=0^+$ and $1^+$, which are chiral partners of $D_s$ and $D_s^*$, respectively. 
} 
the chiral states $/$ chiralons
(, which never exist in the conventional scheme based on NRQM), 
in addition to the Pauli-states $/$ Pauli particles.
The purpose of this talk is to review the essential points of the covariant classification scheme and to explain
the physical background for the existence of chiral states.

({\it Lorentz covariance for local particle})\ \ \ For this purpose it is 
instructive to remember the meaning
of Lorentz-covariance in the case of local fields.

For the infinitesimal Lorentz transformation $\Lambda_{\mu\nu}$ of the space time coordinate $X$
the wave function $\Phi (X)$ of the particle field transforms by the $S(\Lambda )$ as,
\begin{eqnarray}
X_\mu^\prime &=& \Lambda_{\mu\nu} X_\nu \equiv (\delta_{\mu\nu} + \epsilon_{\mu\nu})X_\nu \ , \label{eq1}\\
\Phi (X) &\rightarrow & \Phi^\prime (X^\prime ) = S(\Lambda ) \Phi (X)\  , \label{eq2}\\
   && S(\Lambda ) \equiv (1+\frac{i}{2}\epsilon_{\mu\nu}\Sigma_{\mu\nu})\ . \label{eq3}
\end{eqnarray}
The generators for rotation $J_i$ and for boost $K_i$ are defined in terms of $\Sigma_{\mu\nu}$, respectively, as
\begin{eqnarray}
J_i &\equiv & \frac{1}{2}\varepsilon_{ijk}\Sigma_{jk}\ ,\ \ K_i = i \Sigma_{i4}\ .
\label{eq4}
\end{eqnarray}
In the case of Dirac spinors with spin $J=1/2$, the transformation operators$/$generators for rotation and boost are
given, respectively, as 
\begin{eqnarray}
S_R(\theta_i) &= & e^{-i \mbox{\boldmath$\theta$}\cdot\mbox{\boldmath$J$}}\ , \ \ 
  J_i \equiv \frac{1}{2}\sigma_i\bigotimes\rho_0\ ;  \label{eq5}\\
S_B(\mbox{\boldmath$P$}) &= & e^{-i \mbox{\boldmath$b$}\cdot\mbox{\boldmath$K$}}\ , \ \ 
  (\mbox{\boldmath$b$}\equiv \hat{\mbox{\boldmath$v$}}\ {\rm cosh}^{-1} v_0),\ \ \ 
  K_i \equiv \frac{i}{2}\rho_1\bigotimes\sigma_i\ ;  \label{eq6}
\end{eqnarray}
where $\theta_i$ are rotation angles around the $i$-axes, $v_\mu \equiv P_\mu /M$
is the four-velocity of relevant particles, and $\rho_i$ and $\sigma_i$ is the $2\times 2$ Pauli-matrices
representing the $4\times 4$ Dirac matrices as $\gamma \equiv \rho \bigotimes \sigma$.
The $\rho$ and $\sigma$ concern, respectively, with the transformation 
between the upper-two components of Dirac-spinor $\psi$ (as a whole-entitiy)
and the lower-two, and with the transformation within the respective two-components
as,
\begin{eqnarray}
\psi & = & \left( \begin{array}{c} \varphi_1\\ \varphi_2 \\ {\rm - - - -}\\ \chi_1\\  \chi_2 \\  \end{array} \right)
\begin{array}{l}  \nwarrow \sigma \\  \swarrow \\ {\rm - - -}\\ \nwarrow\sigma\\ \swarrow \\ \end{array}
\begin{array}{r}  \nwarrow  \\  |  \\ {\rm - - -}\\  |  \\ \swarrow \\ \end{array}\ \rho\ .
\label{eq7}
\end{eqnarray}
By inspecting the above formulas, noting that the $\rho$ is contained only in the booster $K_i$,
we see that the boosting transformation $S_B$ plays its role in the moving frame, while becomes
identity in the rest-frame of relevant particle as $S_B(\mbox{\boldmath$P$}\rightarrow 0)\rightarrow 1$.
This means physically that, all the four components are necessary for its covariant description, although
only the two-components are enough for description of its spin $1/2$ character.

\section{Covariant framework for describing composite hadrons}

({\it Attributes and wave function of hadrons})\ \ \ Our relevant composite hadrons have, as
their indispensable attributes, definite mass and spin, definite Lorentz-transformation property, and
definite quark-composite structures.
Accordingly, in any covariant framework, $\bigcirc\hspace{-0.3cm}1$ the hadron wave function (WF)
must satisfy the Klein-Gordon equation, $\bigcirc\hspace{-0.3cm}2$ the explicit form of generators for its
Lorentz-transformation is to be given, and  $\bigcirc\hspace{-0.3cm}3$ the WF has the spinor-flavor
indices $A(\alpha ,\ a)$ $(\alpha =1\sim 4 \ (a)$ denoting Dirac (flavor) index).
Thus, we set up the WF for mesons and baryons, respectively, as   
\begin{eqnarray}
{\rm Meson} & : & \Phi_A{}^B(x,y) \sim \psi_A(x) \bar\psi^B(y)\ , \ \nonumber\\
{\rm Baryon} & : & \Phi_{A_1A_2A_3} \sim \psi_{A_1}(x_1)\psi_{A_2}(x_2)\psi_{A_3}(x_3)\ , \ \ \ \ 
\label{eq8}
\end{eqnarray}
and assume that they are tensors in the $\tilde U(12)_{SF}\bigotimes O(3,1)$ space (the $O(3,1)$
being the Lorentz space for the space-time coordinates of constituent quarks).

({\it Klein-Gordon equation and mass term})\ \ \ We start 
from the Yukawa-type Klein Gordon equation as a basic wave equation\cite{rf6}.
\begin{eqnarray}
[  \partial^2/\partial X_\mu^2   &-& 
 {\cal M}^2 ( r_\mu ,\partial /\partial r_\mu ;
\partial /i \partial X_\mu  ) ] \Phi (X,r,\cdots ) =0\ ,\ \ 
\label{eq9}
\end{eqnarray}
where $X_\mu (r$ for mesons, $r_1,r_2$ for baryons) are the center of 
mass (relative) coordinates of hadron systems. 
It is here to be noted that the Klein-Gordon equation corresponds to Einstein Relation.
Here the Klein-Gordon operator, both the $\partial^2/\partial X_\mu^2$ and 
the squared mass operator ${\cal M}^2$(consisting of the part of kinetic motion
on relative space-time freedom and of the confining-force part 
${\cal M}^2_{\rm conf.}$) 
is assumed to be Lorentz-scalar and $A,(B)$-independent, leading in the case of 
light-quark hadrons to the squared mass spectra with the $\tilde U(12)$ 
symmetry and also with the chiral symmetry. As its concrete 
model we apply the covariant oscillator in COQM, leading to the 
straightly-rising Regge trajectories. The effects due to perturbative 
QCD and other possible effects are neglected in the symmetric limit. 
The total WF are separated 
into the positive (negative)-frequency parts 
concerning the CM plane-wave motion and expanded in terms of eigen-states 
of the squared-mass operator, $\psi_N^{(\pm )}$ satisfying 
${\cal M}^2 \psi_N^{(\pm )} = M_N^2 \psi_N^{(\pm )}$, as   
\begin{eqnarray}
\Phi (X,r,\cdots) &=& \sum_N \sum_{{\bf P}_N}   
\left[  e^{iP_N\cdot X} \psi_N^{(+)} (P_N,r,\cdots )
     +e^{-iP_N\cdot X} \psi_N^{(-)} (P_N,r,\cdots ) \right] \ .\ \ \ \ \ 
\label{eq10}
\end{eqnarray}

({\it Expansion of WF on bases of BW-spinor $\bigotimes$ Oscillator})\ \ \ 
In order to reproduce the success of the conventional NR-classification scheme with approximate LS-symmetry,
we will describe the internal WF of 
relativistic composite hadrons with 
a definite mass and a definite total 
spin $(J=L+S)$, which are tensors in the $\tilde U(4)_{\rm D.S.}\times O(3,1)$ 
space, by expanding them in terms of manifestly covariant bases of complete set, 
being a direct product of eigen-functions in the respective sub-space.
\begin{eqnarray}
\psi_{J,\alpha\cdots}^{\ \ \ \ \ \beta\cdots} (P_N,r,\cdots)
   &=& \sum_{i, j} c_{ij}^J\ W_\alpha^{(i)\beta} (P_N) O^{(j)}(P_N, r\cdots )\ .  
\label{eq11}
\end{eqnarray}
We choose the BW spinors and the covariant oscillator functions (with a 
definite metric type) as the respective bases.

({\it Spinor\ WF})\ \ \ \ The internal WF is, concerning the 
spinor freedom, expanded in terms of complete set of 
multi-Dirac spinors, Bargmann-Wigner (BW) spinors. 
The BW spinors are defined as solutions of the relevant local Klein-Gordon equation:
\begin{eqnarray}
(  \partial^2/\partial X_\mu^2 &-& M^2  ) 
  W_{\alpha\cdots}^{\ \ \beta\cdots} (X) = 0
\label{eq12}\\  
W_{\alpha\cdots}^{\ \ \beta\cdots} (X) & \equiv & 
\sum_{{\bf P}}
(e^{iPX}W_{\alpha\cdots}^{(+)\beta\cdots}(P)
+ e^{-iPX}W_{\alpha\cdots}^{(-)\beta\cdots}(P)  ) .               
\label{eq13}
\end{eqnarray}

({\it Space-time\ WF})\ \ \ The internal WF is, concerning the 
relative space-time freedom, expanded in terms of the complete set 
of covariant oscillator eigen-functions; where, by applying a 
Lorentz-invariant subsidiary condition\cite{rf7} to 
``freeze" the relative-time freedom, 
\begin{eqnarray}
\langle P_\mu r_\mu \rangle = \langle P_\mu p_\mu \rangle =0
  & \stackrel{{\bf P}=0}{\rightarrow} &  O(3,1) \approx O(3)\ .\ \ \ \  
\label{eq14}
\end{eqnarray}
The original symmetry $O(3,1)$ is reduced into the 
non-relativistic $O(3)$ symmetry.

Here it is noteworthy that the above choice of bases is desirable 
from the phenomenological facts  
i) that the constituent quark inside of hadrons behaves like a free 
Dirac-particle\footnote{
The BW-spinors with total hadron momentum $P_\mu$ and $M$ are easily
shown to be equivalent to the product of free Dirac spinors of the 
respective constituent ``exciton-quarks" with momentum 
$p_{N,\mu}^{(i)}\equiv \kappa^{(i)}P_{N,\mu}$ and mass 
$m_N^{(i)}\equiv \kappa^{(i)}M_N\ \ (\sum_i \kappa^{(i)}=1)$.
See the reference\cite{rf3}.
} (implied by BW-spinors) and  
ii) that in the global structure of hadron spectra (Regge trajectory 
and so on) is well described by the corresponding oscillator potential. \\

({\it Spinor WF and chiral states})\ \ \ 
In order to make clear the physical background for the chiral states we describe the
spinor WF of the ground states of mesons and baryons, neglecting the internal space-time variables.
First we define the Dirac spinor for quarks $W_q$ and its Pauli-conjugate for anti-quarks $\bar W_{\bar q}$
as BW spinors with single index, respectively, by   
\begin{eqnarray}
{\rm Dirac\ spinor}\ \ \ 
\psi_{q,\alpha} (X) &=& \sum_{{\bf P},r} [ e^{iPX} W_{q,\alpha}^{(+)}(P) 
   + e^{-iPX} W_{q,\alpha}^{(-)}(P) ]  \nonumber\\
   & &W_{q,\alpha}^{(+)}(P) = u_\alpha (P),
       \ \ \ W_{q,\alpha}^{(-)}(P) = u_\alpha (-P).  \ \ \ \ \ \ 
\label{eq15}\\
\bar\psi_{\bar q,\alpha} (X) &=& \sum_{{\bf P},s} [ e^{iPX} \bar W_{\bar q,\alpha}^{(+)}(P) 
   + e^{-iPX} \bar W_{\bar q,\alpha}^{(-)}(P) ] \nonumber\\
   & & \bar W_{\bar q,\alpha}^{(+)}(P) = \bar v_{\bar q,\alpha} (P),
         \ \ \ \bar W_{\bar q,\alpha}^{(-)}(P) = \bar v_{\bar q,\alpha} (-P).\ \ \ \ \ \   
\label{eq16}
\end{eqnarray}
They take the following form at the hadron rest frame as
\begin{eqnarray}
\begin{array}{c}
W({\bf P=0}); \\  \ \ \ \\ \end{array} 
  &\ \ &
 \begin{array}{c@{}c@{}cccc@{}c@{}cc} 
 W_{q,r}^{(+)} &=& \left( \begin{array}{c} \chi_r \\ 0 \end{array} \right) & \rho_3=+ & ,
    &  W_{q,r}^{(-)} &=& \left( \begin{array}{c} 0\\  \chi_r  \end{array} \right) & \rho_3=-\\ 
 \bar W_{\bar q,s}^{(+)} &=& (0,\bar\chi_s) & \bar\rho_3=+ & , 
    & \bar W_{\bar q,s}^{(-)} &=& (\bar\chi_s, 0) & \bar\rho_3=- 
\end{array}\ \ \ \ \ \ \ \ 
\label{eq17}
\end{eqnarray}
It is to be noted that all Dirac spinors with positive and negative values of 
$\rho_3(\bar\rho_3)$ spin for quarks (anti-quarks)
are required as members of complete set of expansion bases inside of hadrons.
The spinor WF for ground state mesons are given by bi-Dirac spinors as BW spinors with two indices as  
\begin{eqnarray}
{\rm Meson\ spinor}\ \ \ \ \ &\ \  &  
  W_\alpha^\beta (P)\  =\  W_q(P)_\alpha\ \  \bar W_{\bar q}(P)^\beta   \nonumber\\
   W_\alpha{}^\beta ({\bf P=0}) &;\ & (\rho_3,\bar\rho_3) = \  (\ +,\ +\ ):\ {\rm boosted\ Pauli\ states}\ \ \   
\label{eq18}\\
    & & \ \ (\ +,\ -\ ),\ (\ -,\ +\ ),\ (\ -,\ -\ ):\ {\rm ``Chiral\ States"}  . 
\nonumber
\end{eqnarray}
The spinor WF for ground states of baryons and anti-baryons are given similarly by
tri-Dirac spinors as BW spinors with three indices as
\begin{eqnarray}
{\rm Baryon\ spinor}\ \ \ \ & &
  W_{\alpha\beta\gamma}^{(B)} (P)\ =\  W_{q,\alpha}(P) \  W_{q,\beta}(P)\ W_{q,\gamma}(P)   \nonumber\\
   W_{\alpha\beta\gamma}({\bf P=0}); &\ \  & (\rho_3^{(1)},\rho_3^{(2)},\rho_3^{(3)}) =  \ 
    (\ +, \ +, \ + \ ):\ {\rm boosted\ Pauli\ states}\ \ \ \ \ \ \ \ \ \ \   
\label{eq19}\\
    & & \ \ \ \ \ \ \ \ (\ +, \ +,\ -\  ),\ (\ +,\ -,\ -\ ):\ {\rm ``Chiral\ States"}\nonumber\\
   & & \nonumber\\
    & &  W_{\alpha\beta\gamma}^{(\bar B)} (P)\ =\   
         W^{(B)} \{\  W_{q,\alpha} \rightarrow  W_{\bar q,\alpha}\  \} \ .
\label{eq20}
\end{eqnarray}
The meson WF with $(\rho_3,\bar\rho_3)=(+,+)$ and the baryon WF with 
$(\rho_3^{(1)},\rho_3^{(2)},\rho_3^{(3)})=(+,+,+)$ are the multi-boosted Pauli spinors which 
reduce to the multi-NR Pauli spinors at the rest frame, while the meson and baryon WF with the other values of
$(\rho_3,\bar\rho_3)$ and of $(\rho_3^{(1)},\rho_3^{(2)},\rho_3^{(3)})$, respectively, describe the chiral states
of mesons and baryons, which newly appear in the covariant classification scheme. 

Here it is to be noted that the intrinsic spin-freedom for totality 
of BW spinors of $q\bar q$ mesons is $4\times 4=16$, four times of 
$2\times 2=4$ for boosted Pauli spinors. 
For $qqq$ baryons $G(P)$ and $F(P)$, which 
include one and two negative $\rho_3$ Dirac spinors, respectively, 
appear in addition to the conventional $E(P)$ (with all positive 
$\rho_3$ Dirac spinors), boosted multi-Pauli spinor.
It is also to be noted that, although the BW equation with a definite 
mass itself is not $\tilde U(4)_{DS}$ symmetric, the Klein-Gordon 
equation with a definite mass-squared is generally $\tilde U(4)_{DS}$ symmetric.

({\it Covariant quark representation of hadrons and rule for chiral transformation})\ \ \ 
Any transformation rule for composite hadrons are able to be derived from that for constituent quarks
in the covariant classification scheme. An interesting and useful example is that for chiral transformation:
\begin{eqnarray}
{\rm (Chiral\ Transf.)}\ \ \ \ 
\psi_q &\rightarrow& \psi_q^\prime = e^{i\beta\gamma_5} \psi_q\ , \label{eq201}\\
{\rm Baryon};  W_{\alpha_1\alpha_2\alpha_3}
     &\rightarrow&  W_{\alpha_1\alpha_2\alpha_3}^\prime 
    = \left[ \Pi_{i=1}^3 (e^{i\beta\gamma_5^{(i)}}) W\right]_{\alpha_1\alpha_2\alpha_3}\ , \nonumber\\
{\rm Meson};  W_{\alpha}^\beta &\rightarrow&  W_{\alpha}^{\prime\beta} 
    = \left[  e^{i\beta\gamma_5}  W  e^{i\beta\gamma_5}  \right]_{\alpha}{}^\beta\ . \nonumber\\
\end{eqnarray}
In order to see the physical meaning of the chiral transformation of hadrons we note the results of operating the generators of 
chiral transformation on the constituent quarks inside of hadrons.
\begin{eqnarray}
u_\pm (P)  &\rightarrow& u_{(\pm)}^\prime (P) = -\gamma_5 u(P)_{(\pm )} = u(-P)_{(\mp)}\  ,  \nonumber\\
 v_\pm (P) &\rightarrow& v_{(\pm)}^\prime (P) = -\gamma_5 v(P)_{(\pm )} = v(-P)_{(\mp)}\  .  
\label{eq21}
\end{eqnarray}
Noting that $P_\mu \equiv ({\bf P}, iE_{\bf P}=i\sqrt{{\bf P}^2+M^2})$ the transformation changes the positive 
$\rho_3\ (\bar\rho_3)$-quark (anti-quark) spinors into the negative ones, and vice versa. This implies 
that the chiral transformation transforms the members of complete set of BW-spinors for 
each other. Accordingly, if ${\cal M}^2$ operator is 
a Lorentz-scalar and independent of Dirac indices, the hadron squared-mass 
spectra have effectively the chiral symmetry, in addition to the 
$\tilde U(4)_{\rm D.S.}$ symmetry and 
$\tilde U(12)_{\rm SF}$ symmetry(, 
including the flavor freedom in the case of 
light quark system).
Here it is to be stressed that the intention in this work is not to treat 
a dynamical problem from a conventional composite picture, but is to 
propose a \underline{kinematical\ framework} for describing composite 
hadrons covariantly. The validity of the above assumption is checked 
only by comparing its predictions with experimental and phenomenological
facts.

\section{Level Structure of Mesons and Baryons}

({\it Symmetry of ground states})\ \ \ In our scheme 
hadrons are generally classified as the members of multiplet in the
$\tilde U_{SF}(12)\times O(3,1)$ scheme.
The light-quark ground state mesons and baryons are assigned to the representations
$(\underline{\bf 12}\times \underline{\bf 12}^*)=\underline{\bf 144}$ and 
$(\underline{\bf 12}\times\underline{\bf 12}\times 
\underline{\bf 12} )_{\rm Symm}=\underline{\bf 364}$ of the $\tilde U(12)_{SF}$ symmetry, respectively.
(See, Tables \ref{tab2} and \ref{tab3}.)
The numbers of freedom of spin-flavor WF in NRQM are 
${\bf 6}\times {\bf 6}^*=\underline{\bf 36}$ for mesons and 
$({\bf 6}\times {\bf 6}\times {\bf 6})_{\rm Symm.}
=\underline{\bf 56}$ for baryons (and \underline{\bf 56}$^*$ for antibaryons): 
These numbers in COQM are extended to 
\underline{\bf 144} for mesons, and 
 $\underline{\bf 364}=\underline{\bf 182}$ (for baryons)
$+\underline{\bf 182}$ (for anti-baryons) for baryon-and-antibaryons, respectively.

\begin{table}
\begin{center}
\begin{tabular}{l@{}c@{}l}
Mesons: &   $\begin{array}{ccc}  ({\bf 12} & \times & {\bf 12})\\  \end{array}$   & $={\bf 144}$\hspace{2cm}{\bf Chiral States} \\
             &   $\begin{array}{ccc}   & P_s^{(N)}  &  V_\mu^{(N)}\\
                        J^{PC}  &  0^{-+}  &  1^{--} \\ \end{array}$    
          &  $\begin{array}{|cccccc|}\hline  Ps^{(E)} & V_\mu^{(E)} & S^{(N)} & A_\mu^{(N)} & S^{(E)} & A_\mu^{(N)} \\
                       0^{-+} & 1^{--} & 0^{++} & 1^{++} & \underline{0^{+-}} & 1^{+-}\\  \hline   \end{array}$
\end{tabular}
\end{center}
\caption{Quantum numbers of ground-state meson multiplet in $\tilde U(12)_{SF}$ symmetry}
\label{tab2}
\end{table}

\begin{table}
\begin{center}
\begin{tabular}{lr@{}l}
Baryons: &   $({\bf 12} \times {\bf 12} \times {\bf 12})_{Sym}$   & $={\bf 364}={\bf 182}_B+{\bf 182}_{\bar B}$ \\
       &   {\bf 182}---- & $\left[ \begin{array}{l}   
  \begin{array}{cccc}  {\bf 56} & \ \  &  \Delta_{3/2}^{\bigcirc\hspace{-0.2cm}+}  & N_{1/2}^{\bigcirc\hspace{-0.2cm}+} \\  \end{array}\\
  \begin{array}{|ccccc|} \hline  {\bf 70} & \ \ & \Delta_{1/2}^{\bigcirc\hspace{-0.2cm}-(\bigcirc\hspace{-0.2cm}+)}
           & N_{3/2}^{\bigcirc\hspace{-0.2cm}-(\bigcirc\hspace{-0.2cm}+)}  &  N_{1/2}^{\bigcirc\hspace{-0.2cm}-(\bigcirc\hspace{-0.2cm}+)}     \\  
 {\bf 56}^\prime & \ \ & \Delta_{3/2}^{\bigcirc\hspace{-0.2cm}+(\bigcirc\hspace{-0.2cm}-)}
           & N_{1/2}^{\bigcirc\hspace{-0.2cm}+(\bigcirc\hspace{-0.2cm}-)}  &       \\  
  \hline   \end{array}\\
\end{array}\right.$    \\
  &  &   \ \ \ \ \ \ \ \ \ \ \ \ \hspace{1cm} {\bf Chiral States}\\
\end{tabular}
\end{center}
\caption{Quantum numbers of ground-state baryon multiplet in $\tilde U(12)_{SF}$ symmetry}
\label{tab3}
\end{table}

Inclusion of heavy quarks\footnote{
Concerning the heavy quarks, we take only the boosted-Pauli spinors, $\tilde U(4)_{P.S.}$
as physical ones out of total members of $\tilde U(4)_{D.S.}$, 
because that for them the chiral symmetry
is not valid. The $\tilde U(4)_{P.S.}$ symmetry becomes 
a two dimensional unitary symmetry at the rest frame $U(2)_{\rm stat}$.
} 
is straightforward: The WF of general 
$q$ and$/$or $Q$ hadrons become tensors in $O(3,1)\bigotimes 
[\tilde U(4)_{\rm D.S.}\bigotimes SU(3)_{\rm F}]_q
\bigotimes [\tilde U(4)_{\rm P.S.}\bigotimes U(1)_{\rm F}]_Q$.

({\it Level\ structure\ of\ ground\ state\ 
mesons})\cite{rf3}\ \ \ \ In Table \ref{tab4} we have summarized 
the properties of ground state mesons in the light and$/$or heavy 
quark systems. It is remarkable that there appear new multiplets 
of the scalar and axial-vector mesons, chiralons, in the $q$-$\bar Q$ and 
$Q$-$\bar q$ systems and that in the $q$-$\bar q$ systems the 
two sets (Normal and Extra) of pseudo-scalar and of vector meson 
nonets exist. The $\pi$ nonet ($\rho$ nonet) is assigned to the
$P_s^{(N)}\ (V_\mu^{(N)})$ state. 

\begin{table}
\begin{center}
\begin{tabular}{l|c|l|lcl}
\hline
   & mass & Approx. Symm. & Spin WF & $SU(3)$ & Meson Type \\
\hline
$Q\bar Q$ & $m_Q+m_{\bar Q}$ & $LS$ symm. & $u_Q(P)\bar v^{\bar Q}(P)$
          & $\underline{1}$ & $P_s,\ V_\mu$  \\
\hline
$q\bar Q$ & $m_q+m_{\bar Q}$ & $q$-Chiral Symm. & $u_q(P)\bar v^{\bar Q}(P)$
          & $\underline{3}$ & $P_s,\ V_\mu$  \\
               &                              & $\bar Q$-Heavy Q. Symm. 
          & $u_q(-P)\bar v^{\bar Q}(P)$ & $\underline{3}$ & $S,\ A_\mu$  \\
$Q\bar q$ & $m_Q+m_{\bar q}$ & $\bar q$-Chiral Symm.
          & $u_Q(P)\bar v^{\bar q}(P)$
          & $\underline{3}^*$ & $P_s,\ V_\mu$  \\
               &                              & $Q$-Heavy Q. Symm. 
          & $u_Q(P)\bar v^{\bar q}(-P)$ & $\underline{3}^*$ & $S,\ A_\mu$  \\
\hline
$q\bar q$ & $m_q+m_{\bar q}$ & Chiral Symm.
     & $\frac{1}{\sqrt{2}}(u(P)\bar v(P)\pm u(-P)\bar v(-P) )$
     & \underline{\bf 9} & $P_s^{(N,E)},\ V_\mu^{(N,E)}$  \\
              &                             &   
     & $\frac{1}{\sqrt{2}}(u(P)\bar v(-P)\pm u(-P)\bar v(P) )$
     & \underline{\bf 9}   &  $S^{(N,E)},\ A_\mu^{(N,E)}$  \\
\hline
\end{tabular}
\end{center}
\caption{Level structure and the spinor wave function of ground-state mesons}
\label{tab4}
\end{table}

({\it Level\ structure\ of\ mesons\ 
in\ general})\cite{rf3}\ \ \ \ The global mass spectra of the 
ground and excited state mesons are given by
\begin{eqnarray} 
M_N^2 & = & M_0^2+N\Omega = (m_N^{(1)} + m_N^{(2)})^2 .
\label{eq41}
\end{eqnarray}
Their quantum numbers are given in Table \ref{tab5}. Here it is 
to be noted that some chiralons have the ``exotic" quantum numbers
from the conventional NRQM viewpoint.

\begin{table}
\begin{center}
\begin{tabular}{c@{}||@{}c} \hline
$\begin{array}{c|c|c|c}
(q\bar q)  & P & C & N \\
\hline
\{P_s^{(N)},V_\mu^{(N)}  \} \bigotimes \{ L,N \} & 
(-1)^{L+1} & (-1)^{L+S} & {\rm all}\\
\{P_s^{(E)},V_\mu^{(E)}  \} \bigotimes \{ L,N \} & 
(-1)^{L+1} & (-1)^{L+S} & 0,1\\
\end{array}$
 &  
$\begin{array}{c|c|c}
 (q\bar Q\  {\rm or}\  Q\bar q) & P & N \\
\hline
 \{ P_s,V_\mu \} \bigotimes \{ L,N \} & (-1)^{L+1} & {\rm all}\\ 
 \{ S,A_\mu \} \bigotimes \{ L,N \} & (-1)^{L} & 0,1\\
\hline
\end{array}$ \\
$\begin{array}{c|c|c|c}
\{ S^{(N)},A_\mu^{(N)}  \} \bigotimes \{ L,N \} & 
(-1)^{L\ \ \ } & (-1)^{L} & 0,1\\
\{ S^{(E)},A_\mu^{(E)}  \} \bigotimes \{ L,N \} & 
(-1)^{L\ \ } & (-1)^{L+1} & 0,1\\
\end{array}$ 
 & 
$\begin{array}{c|c|c}\hline 
 (Q\bar Q) & P & N \\
\hline
 \{ P_s,V_\mu \} \bigotimes \{ L,N \} & (-1)^{L+1} & {\rm all} \\
\end{array}$ \\
\hline
\end{tabular}
\end{center}
\caption{Level structure of Mesons:
We are able to infer\cite{rf8} that the chiral symmetry concerning the light quarks is valid (still effective)
for the ground (first excited) state of $n\bar n$ and $n\bar Q$ meson systems, while the
symmetry will prove invalid from the $N$-th ($N\geq 2$) excited hadrons.}
\label{tab5}
\end{table}

({\it Level\ structure\ of\ baryons})\ \ \ \  
The baryon WF in Eq.~(\ref{eq19}) should be full-symmetric (except 
for the color freedom) under exchange of constituent quarks: The 
full-symmetric total WF in the extended scheme is obtained, in 
the three ways, as a product of the sub-space $\rho$, $\sigma$ and $F$ WF with
respective symmetric properties. (As for details, see ref. \citen{rf9}.)

Here it is remarkable that there appear chiralons in the ground states.
That is, the extra positive parity $\underline{\bf 56}^\prime$-multiplet of the 
static $SU(6)$ and the extra negative parity \underline{\bf 70}-multiplet 
of the $SU(6)$ in the low mass region. It is also to be noted that the 
chiralons in the first excited states are expected to exist. The above 
consideration on the light-quark baryons are extended directly to the 
general light and$/$or heavy quark baryon systems: The chiralons are 
expected to exist also in the $qqQ$ and $qQQ$-baryons, while no chiralons 
in the $QQQ$ system.

\section{Remarks on Interaction among Hadrons}

({\it Chiral\ symmetric\ spectator})\ \ \ \ 
In treating interactions between hadrons resorting on their quark-composite structure,
the effective interaction verices are generally written as a product of the two parts,
the one concerning the fundamental quark-interactions and the other concerning the overlap
of spectator quarks. The overlapping spectator-interaction, which is considered to be due to
QCD, should be chiral symmetric in the ideal limit of neglecting the effects of spontaneous breaking. 
However, the bilinear scalar-covariant between Dirac spinor
$\psi$ and their Pauli-conjugate $\bar\psi\equiv \psi^\dagger \gamma_4$ ($\bar\psi\ 1\  \psi$)
violates maximally the chiral symmetry, leading to the non-chiral symmetric effective hadron
interaction, even in the case of symmetric fundamental quark interactions. Correspoindingly
we define the ``unitary WF" (its conjugate) of mesons and of baryons, respectively, (revising the WF (\ref{eq8}))
by
\begin{eqnarray}
{\rm Meson} & : & \Phi_{U,A}{}^B \equiv (\Phi\bar\gamma_4 )_A{}^B\sim \psi_A\psi^{\dagger B} , 
\ \ \bar\Phi_{U,B}{}^A \equiv (\bar\Phi\bar\gamma_4 )_B{}^A\sim \psi_B\psi^{\dagger A} 
\label{eq42}\\
  & &  (\bar\Phi \equiv \gamma_4 \Phi^\dagger \gamma_4,\ \ \  \bar\gamma_4=-i v\cdot\gamma ,\ \ 
 v_\mu \equiv P_\mu /M \  )\ .  
\nonumber\\
{\rm Baryon} & : & \Phi_{U,A_1A_2A_3} \equiv \Phi_{A_1A_2A_3} \sim \psi_{A_1}\psi_{A_2}\psi_{A_3}\ , 
\nonumber\\
   & & \bar\Phi_{U}^{B_1B_2B_3} \equiv (\bar\Phi\Pi_{i=1}^3 \gamma_4^{(i)})^{B_1B_2B_3} 
    \sim \psi^{\dagger B_1} \psi^{\dagger B_2}\psi^{\dagger B_3}
\label{eq43}\\
  & &  (\bar\Phi \equiv  \Phi^\dagger \Pi_{i=1}^3 \gamma_4^{(i)},\ \ \  
\bar\gamma_4^{(i)}=-i v\cdot\gamma^{(i)} )\ .  
\nonumber
\end{eqnarray}
These new WF are so defined as leading to their bilinear scalar-covariant being chiral symmetric and also 
equal to the unitary overlap in the static limit
\begin{eqnarray}
\langle \bar\Phi_U \Phi_U \rangle      
  & \stackrel{\mbox{\boldmath$v=0$}}{\longrightarrow} & \langle \Phi^\dagger \Phi \rangle  \ .
\label{eq44}
\end{eqnarray}
Here it is to be noted that each spectattor overlap-interaction now
leads to an interesting selection rule
({\it $\rho_3$-line rule}). In the static limit
\mbox{\boldmath$(v \rightarrow 0)$}  
$\rho_3$ value conserves along each spectator-quark line.

({\it Chiral-symmetric effective EM-currents})\ \ \ \ 
The electromagnetic (EM) effective hadron interactions should be chiral symmetric in the ideal limit. 
We are able to derive the effective EM hadron currents 
(which are conserved and chiral symmetric), following directly to the 
method of minimal substitution\cite{rf10} in the framework of COQM (with replacement
of $\Phi$ by $\Phi_U$).
That is, in the multi-local Lagrangian density of action $S_I^{EM}$, leading to our basic Klein-Gordon
equation (\ref{eq9}), the derivative on quark coordinates is replaced by the gauge-covariant one.
The results are given as follows: 
\begin{eqnarray}
S_I^{EM} & = & \int \Pi_{i=1}^3 d^4x_i \sum_{i=1}^3 j_{i,\mu}^{EM} (x_1,\cdots ) A_\mu (x_i) 
      \equiv \int d^4X J_\mu^{EM}(X) A_\mu (X),\nonumber\\
  & &  j_{i,\mu}^{EM} (x_1,\cdots ) \propto e_i \langle \bar\Phi_U (x_1,\cdots )
           \stackrel{\leftrightarrow}{\partial_{i,\mu}} \Phi_U(x_1,\cdots ) \rangle\ , 
\label{eq45}
\end{eqnarray}
where only the case of baryons is shown as an example. In the actual application this current should
be revised by taking into account the symmetry breaking effects above mentioned.

\section{Candidates of Chiral States and Concluding Remarks}

({\it Experimental candidates of chiral particles})\ \ \ \ 
In our level-classification scheme a series of new type of multiplets, 
chiralons, are predicted to exist in the ground and 
the first excited states. Presently we can 
give only a few experimental candidates or indications for them:\\
\underline{\em ($q\bar q$-mesons)}\footnote{
A tentative assignment in the covariant classification scheme of all the light-quark mesons with mass 
$\stackrel{<}{\scriptstyle \sim}$ 1.8GeV, reported in PDG, was made rather satisfactorily. (See Ref. \citen{rf18}.)
}\ \ \ \ One of the most important 
candidates is the scalar $\sigma$ nonet to be assigned as $S^{(N)}(^1S_0)$ : 
$[\sigma (600),\ \kappa (900),\ a_0(980),\ f_0(980)]$. The existence of 
$\sigma (600)$ seems to be established\cite{rf1} through the analyses of, 
especially, $\pi\pi$-production processes. 
Some evidences for $\kappa (800$-900) in $K\pi$ scattering phase shift\cite{rf11}
had been given formerly. Its 
firm experimental evidences in the production process\cite{rf12} and through the decay 
process\cite{rf13} were reported recently.\\ 
In our scheme respective two sets of $P_s$- and of $V_\mu$-nonets, to 
be assigned as $P_s^{(N,E)}(^1S_0)$ and $V_\mu^{(N,E)}(^3S_1)$, are 
to exist: The vector mesons\cite{rf14} $\rho (1250)$ and $\omega (1250)$, suspected to exist
for long time,
are naturally able to be assigned as the members of 
$V_\mu^{(E)}(^3S_1)$-nonet;\\
Out of the three established $\eta$, $[\eta (1295),\ \eta (1420),\  
\eta (1460) ]$ at least one extra, plausibly $\eta (1295)$ with the 
lowest mass, may belong to $P_s^{(E)}(^1S_0)$ nonet.\\
Recently the existence of two ``exotic" particles $\pi_1(1400)$ and 
$\pi_1(1600)$ with $J^{PC}=1^{-+}$ and $I=1$, observed\cite{rf15}   
in the $\pi\eta ,\  \rho\pi$ and other channels, is attracting strong 
interests among us.
These exotic particles with a mass around 1.5GeV may be naturally 
assigned as the first excited states $S^{(E)}(^1P_1)$ and 
$A_\mu^{(E)}(^3P_1)$ of the chiralons.\\
(\underline{$q\bar Q$}\ or\ \underline{$Q\bar q$}-mesons)\footnote{
See also the footnote on p.2 .
}\ \ \ \ Recently 
we have shown some indications for existence of the 
following two chiralons\cite{rf16} in $D$- and $B$-meson systems 
obtained through analyses of the $\Upsilon (4S)$ or $Z^0$ decay processes, 
respectively, as \ \ 
$D_1^\chi (2310)$ with $J^P=1^+$ in $D_1^\chi \rightarrow D^* + \pi$ and 
$B_0^\chi (5520)$ with $J^P=0^+$ in $B_0^\chi \rightarrow B + \pi  $.\\   
(\underline{\em $qqq$}-baryons)\ \ \ \  The two facts have been a 
longstanding problem that the Roper resonance $N(1440)_{1/2^+}$ 
is too light to be assigned as radial excitation of $N(939)$ and 
that $\Lambda (1405)_{1/2^-}$ is too light as the $L=1$ excited 
state of $\Lambda (1116)$. In our new scheme these two problems dissolve\cite{rf17} in principle,
because in the ground states there exist the two {\bf 56} of $SU(6)$ with positive parity,
$E({\bf 56}^+)$ and $F({\bf 56}^{\prime +})$, and one {\bf 70} of $SU(6)$ with negative parity, 
$G({\bf 70}^-)$.

({\it Concluding remarks})\ \ \ \ 
We have summarized in this talk the essential points of the covariant level-classification scheme,
which has, we believe, a possibility to solve the serious problem in hadron spectroscopy mentioned in \S 1.
In this connection further investigations, both experimental and theoretical, for chiral states predicted in this scheme,
are urgently required for new development of hadron physics.

\acknowledgements

One of the authors (S.I.) should like to thank deeply to the members of elementary particle
physics group in Nihon University, especially to Professor S.~Y.~Tsai, for their collaboration and support
in holding this symposium.

He wish to take this opportunity to express his sincere gratitude to the late professor O.~Hara and T.~Goto.
He is also grateful to Nihon University(, now retiring from) and 
the colleagues for the past forty-two years.

\end{document}